\documentclass[preprint,showkeys,secnumarabic,amsfonts,showpacs,amsmath,amssymb]{revtex4}
\usepackage[dvips]{color}
\usepackage{array}
\usepackage{rotating}
\usepackage{epsfig}
\usepackage{amsmath}
\usepackage{graphicx}
\begin{document}
\title{ Universe acceleration and fine structure constant variation in BSBM theory}

\author{H. Farajollahi$^{1,2}$}
\email{hosseinf@guilan.ac.ir}
\author{A. Salehi$^{1}$}
\email{a.salehi@guilan.ac.ir}
\affiliation{$^1$Department of Physics, University of Guilan, Rasht, Iran}
\affiliation{$^2$ School of Physics, University of New South Wales, Sydney, NSW, 2052, Australia}

\date{\today}

\begin{abstract}
 \noindent \hspace{0.35cm}
 
 In this work we investigate the utility of using SNe Ia observations in constraining the cosmological parameters in BSBM theory where a scalar field is responsible for both fine structure
constant variation and late time universe acceleration. The model is discussed in the presence of an
exponential self potential for the scalar field. Stability and phase space analysis of the solutions are
studied. The model is tested against observational data for Hubble parameter and quasar absorption spectra. With the best fitted model parameters, the theory predicts a good match with the
experimental results and exhibits fine structure constant variation. The analysis also shows that
for the equation of state parameter, recent universe acceleration and possible phantom crossing in
future is forecasted.

\end{abstract}


\maketitle

\section{introduction}

Possible variations of
fundamental constants which has been the subject of intense speculation and research is pioneered by the work of Dirac in 1937 \cite{dirac}. Among fundamental constants the most observationally sensitive constants is the electromagnetic
fine structure constant, $\alpha=e^2/\hbar c$. Due to the first observational evidence from the quasar absorption spectra the fine
structure constant might change with cosmological time; smaller than its present value by $\frac{\Delta \alpha}{\alpha}\equiv \frac{\alpha -\alpha_0}{\alpha_0}\sim 10^{-5}$ at redshifts in the range $z\sim 1-3$ \cite{webb}-\cite{webb4}. A varying $\alpha$ might be due to the variation of speed of light $c$ \cite{light}-\cite{light2}, while breaks Lorentz invariance, or a varying electron charge $e$ originally proposed by Bekenstein \cite{beken} which preserves local gauge and Lorentz invariance, and is generally covariant. The Bekenstein model has been revived and generalized after the first observational evidence of varying $\alpha$ from the quasar absorption spectra.  The so called BSBM model \cite{bsbm} is one example of this type of models where a dilaton field coupled to the electromagnetic part
of the Lagrangian is responsible for the variation of fine structure constant.

On the other hand, mostly an scalar field introduced into the standard general relativity \cite{Mainini}-\cite{Granda}, coupled to the curvature (for example in Brans-Dick theory \cite {brans}-\cite{Setare}) or matter field (such as in chameleon cosmology \cite{chamel}-\cite{Farajollahi}) represents dark energy (DE) and thus accounts for universe acceleration. Theoretically such a cosmological scalar field could also be coupled to
the electromagnetic field, and hence could drive both the variation of cosmological constants( such as $\alpha$) and universe acceleration. So, one can generalize the BSBM theory and propose a scalar field that plays the role of dark energy for universe acceleration and by its coupling to the electromagnetic field is responsible for fine structure constant variation. So far, the varying
$\alpha$ models driven by a quintessence scalar field or by phantom filed with negative model parameter $\omega$, as in the BSBM model have been extensively investigated in the literature \cite{quintes}-\cite{bsbm4} .

 From observational point of view, any cosmological model, promising to explain the universe acceleration, has to be fitted with the recent observational data from Type Ia Supernovae (SNe Ia) for distance modulus. In addition, for the model, to illustrate fine structure constant variation, it has to be verified by the observational evidence from quasar absorption spectra. To integrate these two disciplines, in this manuscript we begin with the BSBM theory and drive the solutions to the field equations by best fitting the model parameters with the observational data from SNe Ia for distance modulus using chi-squared method. Stability analysis is also performed to determine the best fitted dynamical state of the universe. Finally, we examine the model against the observational data for hubble parameter and also the quasar absorption spectra to verify both $\alpha$ variation and universe acceleration. We also reconstruct the equation of state (EoS) parameter and the scalar field responsible for fine structure constant variation with the best fitted model parameters to reproduce the current universe acceleration.

\section{The Model and observational cosntraints}

In the BSBM theory, the action describing the dynamics of the Universe with a varying-$\alpha$ takes the form
\begin{eqnarray}\label{action}
S=\int d^{4}x\sqrt{-g}(\frac{-1}{16\pi G}R+\frac{\omega}{2}g_{\mu\nu}\partial^\mu\phi\partial^\nu\phi+V(\phi)+{\cal L}_{matter}+{\cal L}_{em}e^{-2\varepsilon\phi})
\end{eqnarray}
where $R$ is Ricci scalar, $G$ is the newtonian constant gravity and  $\omega=\frac{\hbar c}{l^2}$ is a coupling constant determines the strength of coupling between scalar field $\phi$ and photons. The characteristic length scale, $l$, is introduced for dimensional reasons and gives the scale down to which the electric field around a point charge accurately obeys Coulomb force law. From the present experimental constraints, the corresponding energy scale, $\frac{\hbar c}{l}$, has to lie between a few tens of MeV and the Planck scale $\sim 10^{19}$ GeV to avoid conflict with experiment. While in the conventional BSBM theory the coupling scalar function is $e^{-2\phi}$, in this work for the purpose of best fitting and testing analysis of the model we introduce $e^{-2\varepsilon\phi}$ with the parameter $\varepsilon$ as a free parameter. The sign of second term in the action to be positive or negative represents the quintessence or phantom models respectively. The electromagnetic lagrangian is ${\cal L}_{em}=-\frac{1}{4}f_{\mu\nu}f^{\mu\nu}$ where $f_{\mu\nu}$  is the electromagnetic field tensor. The variation of action (\ref{action})  with respect to the metric tensor components in a spatially flat FRW  cosmology
yields the field equations:
\begin{eqnarray}
H^{2}&=&\frac{8\pi G}{3}(\rho_{m}(1+|\zeta|e^{-2\varepsilon\phi})+\rho_r e^{-2\varepsilon\phi}+\frac{\omega}{2}\dot{\phi}^{2}+V(\phi)),\label{fried1}\\
2\dot{H}&+&3H^2=8\pi G(-\frac{1}{3}\rho_r e^{-2\varepsilon\phi}-\frac{\omega}{2}\dot{\phi}^{2}+V(\phi)),\label{fried2}
\end{eqnarray}
where in what follows we put  $8\pi G=c=\hbar=1$. To derive the field equations, we assumed a perfect fluid with $p_{m}=\gamma\rho_{m}$. The energy density $\rho_{m}$ stands for the contribution
from cold dark matter (CDM) to the energy density, so, we can neglect $\rho_r$. In addition, variation of the action (\ref{action}) with respect to scalar field  $\phi$ provides the wave
equation for the scalar field as
\begin{eqnarray}\label{phiequation}
\ddot{\phi}+3H\dot{\phi}+V^{'}=2\frac{\varepsilon|\zeta|}{\omega}\rho_{m}e^{-2\varepsilon\phi}
\end{eqnarray}
where $'$ indicates differentiation with respect to $\phi$.
From equations (\ref{fried1}), (\ref{fried2}) and (\ref{phiequation}), one can easily arrive at the  relation
\begin{eqnarray}\label{conserv}
\dot{\rho_{m}}+3H\rho_{m}=0.
\end{eqnarray}
Now, in the following we constrain the model with the recent observational data of Sne Ia. For the purpose of stability analysis and phase space presentation of the solutions, we first represent the model in terms of new dynamical variables
\begin{eqnarray}\label{conserv1}
\chi=\frac{\sqrt{\rho_{m}}}{\sqrt{3}H},\ \ \xi=\frac{\dot{\phi}}{H},\eta=\frac{\sqrt{V}}{\sqrt{3}H},\theta=\frac{\sqrt{\rho_{m}}}{\sqrt{3}H}e^{-\varepsilon\phi}
\end{eqnarray}
Motivated by stability analysis and cosmological consideration, we shall assume an exponential potential for the scalar field, $V=V_0e^{\beta\phi}$, in which $\beta$ is a dimensionless constants and $V_0$ is a constant with dimensions $[mass]^4$ \cite{barrow1}. The system of coupled second order differential equations (\ref{fried2})-(\ref{phiequation}) in terms of new variables now reduces to the following first order equations,
\begin{eqnarray}
\chi'&=&\frac{\omega\chi\xi^{2}}{4}-\frac{3}{2}\chi\eta^{2}\label{conserv2}\\
\xi'&=&-\beta\eta^{2}+\frac{\omega\xi^{3}}{4}+2\frac{\varepsilon|\zeta|}{\omega}\theta^{2}
-\frac{3}{2}\xi\eta^{2}\label{conserv3}\\
\eta'&=&\frac{\beta\xi\chi}{2}+\frac{3}{2}\eta+\frac{\eta\xi^{2}}{4}-\frac{3}{2}\xi^{3}\label{conserv5}\\
\theta'&=&\frac{\theta\xi^{2}}{4}-\frac{3\theta}{2}\eta^{2}-\varepsilon\xi\theta\label{conserv6}
\end{eqnarray}
where $" ' "$ from now on means derivative with respect to $N = ln (a)$. Also, the Friedmann constraint equation (\ref{fried1}) in terms of the new dynamical variables becomes
\begin{eqnarray}\label{conserv6}
\chi^{2}+\varepsilon|\zeta|\theta^{2}+\eta^{2}+\frac{\omega\xi^{2}}{6}=1
\end{eqnarray}
With the constraint (\ref{conserv6}) we only solve the system of equations (\ref{conserv2})-(\ref{conserv5}). To best fit the model for the parameter $\frac{\varepsilon|\zeta|}{\omega}$ and $\beta$, and the initial conditions $\chi(0)$ and $\xi(0)$, $\eta(0)$ with the most recent observational data, SNe Ia, we employ the $\chi^2$ method. Table I shows the best best-fitted model parameters.

\begin{table}[ht]
\caption{Best-fitted model parameters and initial conditions.} 
\centering 
\begin{tabular}{c c c c c c c c} 
\hline 
 &  $\frac{\varepsilon|\zeta|}{\omega}$  &  $\beta$ \ & $\chi(0)$\ & $\xi(0)$\ & $\eta(0)$\ & $h_0$ \ &
  $\chi^2_{min}$\\ [2ex] 
\hline 
&$0.82 \times 10^{-4}$  & $1.42$ \ & $0.1$\ & $0.6$\ & $1$\ & $0.702$ \ & $547.0858941$ \\
\hline 
\end{tabular}
\label{table:1} 
\end{table}\

Figure 1, show the constraints on the parameters $\frac{\varepsilon|\zeta|}{\omega}$, $\beta$ and at the $68.3\%$, $95.4\%$ and $99.7\%$ confidence level.\\

\begin{tabular*}{2.5 cm}{cc}
\includegraphics[scale=.3]{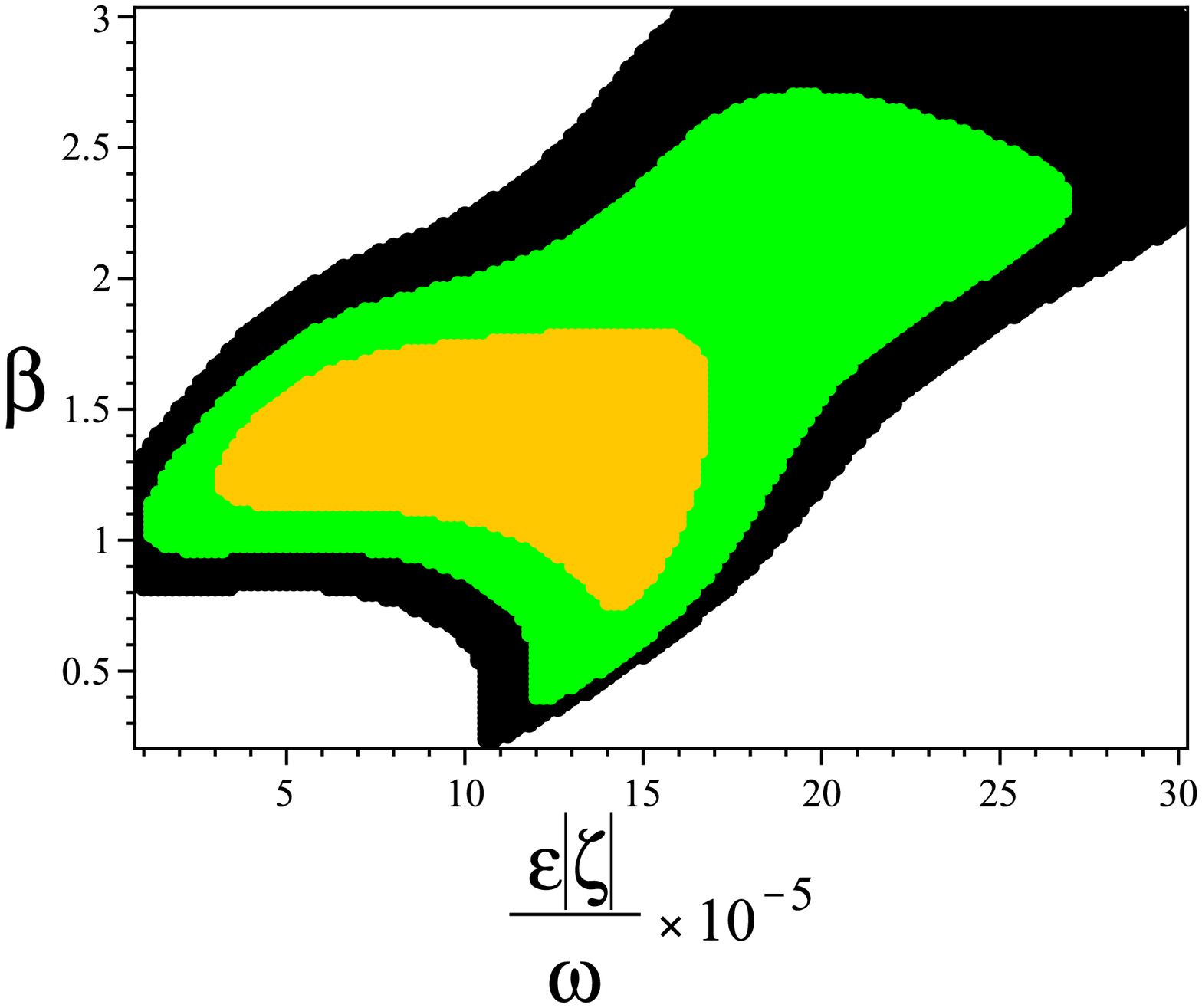}\hspace{0.1 cm}\\
Fig. 1:  The $68.3\%$, $95.4\%$ and $99.7\%$ confidence level for parameters $\frac{\varepsilon|\zeta|}{\omega}$ and $\beta$. \\
\end{tabular*}\\

In the next section we study the phase space analysis of the model with the best fitted model parameters and initial conditions.

\section{Phase space}

The phase space analysis of the model exhibits thirteen critical points. In Table II, the first five of the critical points are unstable and the rest are conditionally stable.

\begin{table}[ht]
\caption{Critical Points coordinates} 
\centering 
\begin{tabular}{c c c c c c c c} 
\hline 
CP &  ($\xi$, $\eta$, $\theta$)\\ [1ex] 
\hline 
$P1$&($0$, $0$, $0$)\  \\
\hline 
$P2_{A,B}$&($\pm\sqrt{6}$, $0$, $0$)\  \\
\hline 
$P3_{A,B}$&($4\varepsilon$, $0$, $\pm\frac{\sqrt{8\varepsilon^{3}-3\varepsilon}}{3\alpha}$)\  \\
\hline 
$P4_{A,B}$&($-\beta$, $\pm\sqrt{\frac{6-\beta^{2}}{6}}$, $0$)\  \\
\hline 
$P5_{A,B}$&($\frac{-3}{\beta}$, $\pm\frac{\sqrt{6}}{2\beta}$, $0$)\  \\
\hline 
$P6_{A,B,C,D}$&($\frac{-3}{\beta+2\varepsilon}$, $\frac{\pm\sqrt{16\varepsilon^{2}+8\beta\varepsilon+6}}{2\beta
+4\varepsilon}$, $\frac{\pm\sqrt{-\alpha\beta^{2}\varepsilon-2\alpha\beta\varepsilon^{2}+3\varepsilon}}{\alpha\beta
+2\alpha\varepsilon}$)\  \\
\hline 
\end{tabular}
\label{table:1} 
\end{table}\

Substituting linear perturbations $\chi'\rightarrow \chi'+\delta \chi'$, $\xi'\rightarrow \xi'+\delta \xi'$ and $\eta'\rightarrow \eta'+\delta \eta'$ about the critical points into the first three independent equations, to the first orders in the perturbations, yields three eigenvalues $\lambda_{i} (i=1..3)$  shown in the following. Unstable points content positive or zero eigenvalues ($ev$) as can be seen explicitly in the first five critical points with the eigenvalues as:
\begin{equation}
 ev1 = \begin{bmatrix}
       0  \\
\frac{3}{2}  \\
-\frac{3}{2}
     \end{bmatrix}
\qquad,
{ev2} = \begin{bmatrix}
    3+\frac{\sqrt{6}}{2}\beta  \\
\frac{3}{2} -\sqrt{6}\varepsilon \\
3
     \end{bmatrix}
     \qquad,
{ev3} = \begin{bmatrix}
    -\frac{3}{2}+4\varepsilon^{2}  \\
8\varepsilon^{2}  \\
\frac{3}{2}+2\beta\varepsilon+4\varepsilon^{2}
     \end{bmatrix}.
\end{equation}
The eigenvalues for the conditional stable critical point have to be negative and are given respectively by
\begin{eqnarray}
 ev4 &=& \begin{bmatrix}
       \frac{\beta^{2}}{2}-3  \\
-3+\beta^{2}  \\
-\frac{3}{2} +\beta\varepsilon+\frac{\beta^{2}}{2}
     \end{bmatrix},
\ \ stable \ for \left\{
   \begin{array}{ll}
     \varepsilon < \frac{3-\beta^{2}}{2\beta},  \ \ 0 < \beta < \sqrt{3}\\ \varepsilon > \frac{3-\beta^{2}}{2\beta}, \ \ -\sqrt{3} < \beta <0 \\ \hbox{$ \varepsilon \ \ \ \ \ \ \ \  \ \ \ \ \ \ \ \ \ \beta=0$}
   \end{array}
 \right.
\\
 ev5 &=& \begin{bmatrix}
       \frac{-3\beta+3\sqrt{-7\beta^{2}+24}}{4\beta}  \\
\frac{-3\beta-3\sqrt{-7\beta^{2}+24}}{4\beta}   \\
\frac{3\varepsilon}{\beta}
     \end{bmatrix},
stable \ for \left\{
   \begin{array}{ll}
     \varepsilon < 0,  \ \  \beta > \sqrt{3}\\ \varepsilon >0, \ \  \beta < -\sqrt{3}  \hbox{ }
   \end{array}
 \right.
\\
 ev6 &=& \begin{bmatrix}
       -\frac{6\varepsilon}{\beta+2\varepsilon}  \\
\frac{-3\beta-12\varepsilon+\sqrt{-63\beta^{2}+216\beta\varepsilon+720\varepsilon^{2}-96\beta^{3}\varepsilon
-384\beta^{2}\varepsilon^{2}-384\beta\varepsilon^{3}+216}}{4\beta+8\varepsilon}  \\
\frac{-3\beta-12\varepsilon-\sqrt{-63\beta^{2}+216\beta\varepsilon+720\varepsilon^{2}-96\beta^{3}\varepsilon
-384\beta^{2}\varepsilon^{2}-384\beta\varepsilon^{3}+216}}{4\beta+8\varepsilon}  \end{bmatrix}
\end{eqnarray}
For the critical points $P4$ and $P5$, Fig. 2 shows the stability region for the parameters $\varepsilon$ and $\beta$.

\begin{tabular*}{2.5 cm}{cc}
\includegraphics[scale=.6]{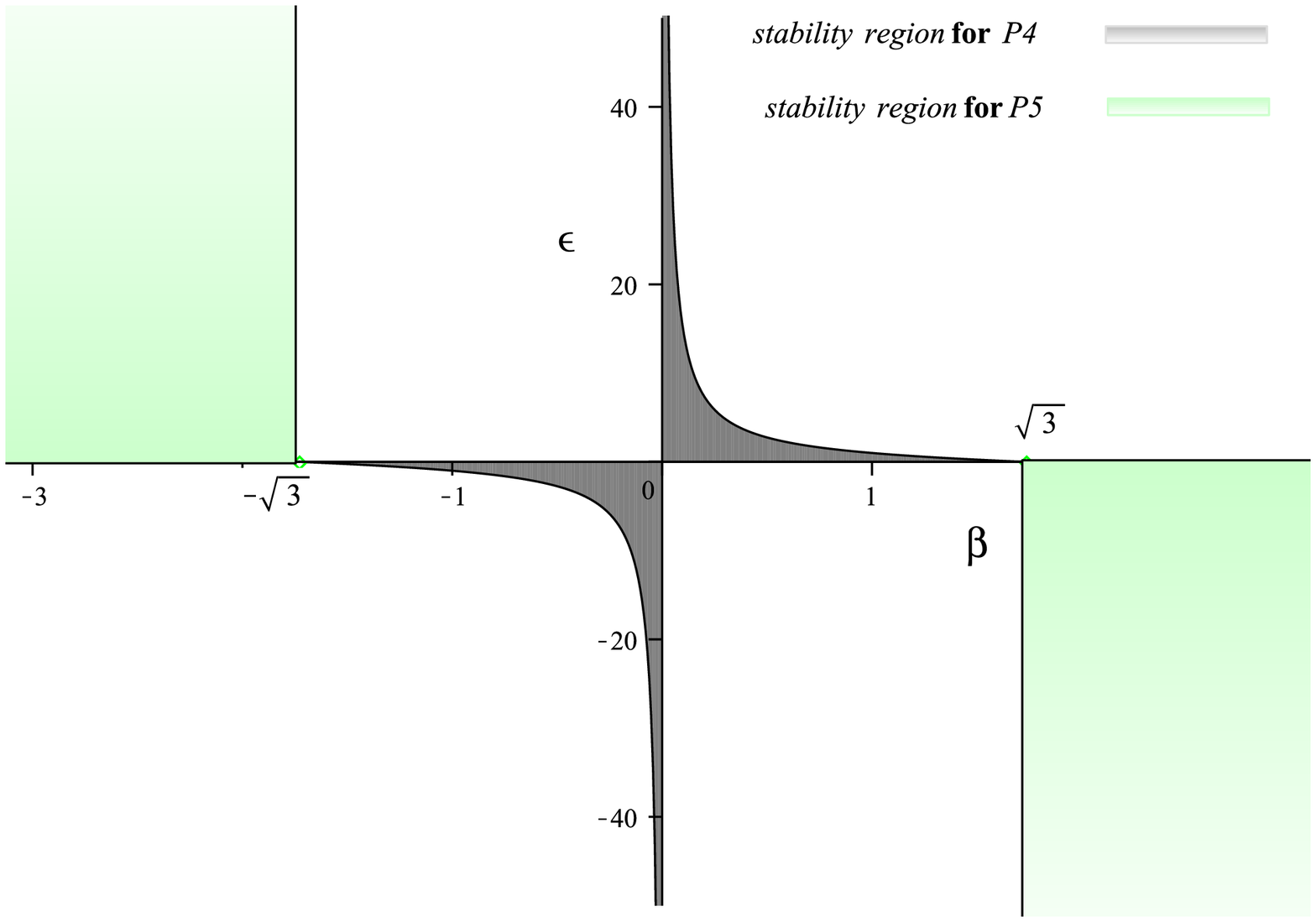}\hspace{0.1 cm}\\
Fig. 2: The stability region for critical points $P4$ and $P5$ \\
\end{tabular*}\\
Figs. 3 and 4 are two examples of the phase space illustrating the behavior of the critical points around which represent the state of universe in our model. The trajectories leaving unstable states $P2_A$ and $P2_B$ may either move towards the stable states $P5_A$, $P5_B$ or $P6_A$, $P6_B$, $P6_C$, $P6_D$, depending on the values of the parameters and initial conditions.

\begin{tabular*}{2.5 cm}{cc}
\includegraphics[scale=.4]{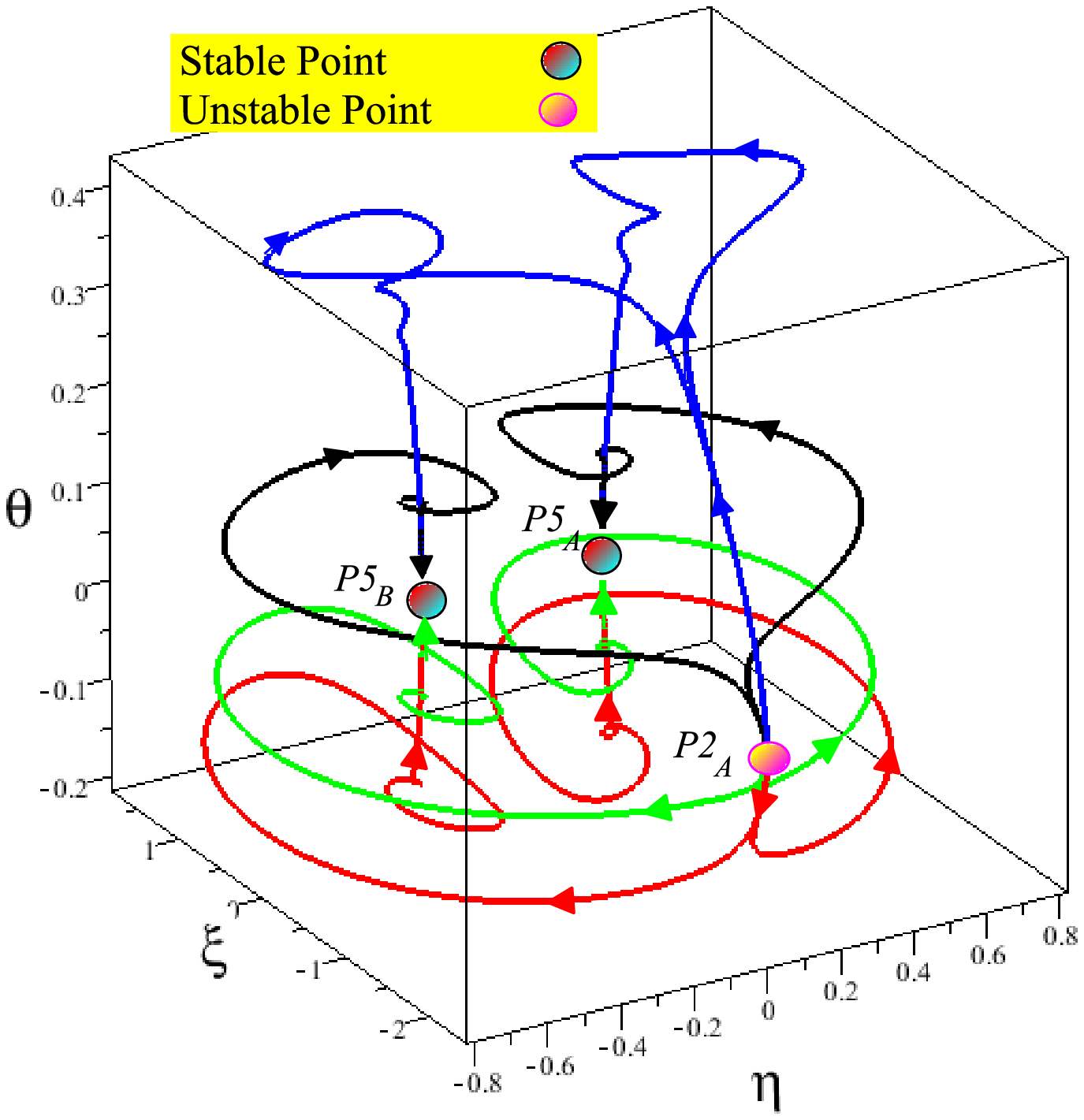}\hspace{0.1 cm}\\
Fig. 3: The phase space graph shows the trajectories leave the unstable critical point $P2_{A}$, \\entering the stable critical points $P5_{AB}$ . \\
\end{tabular*}\\

\begin{tabular*}{2.5 cm}{cc}
\includegraphics[scale=.3]{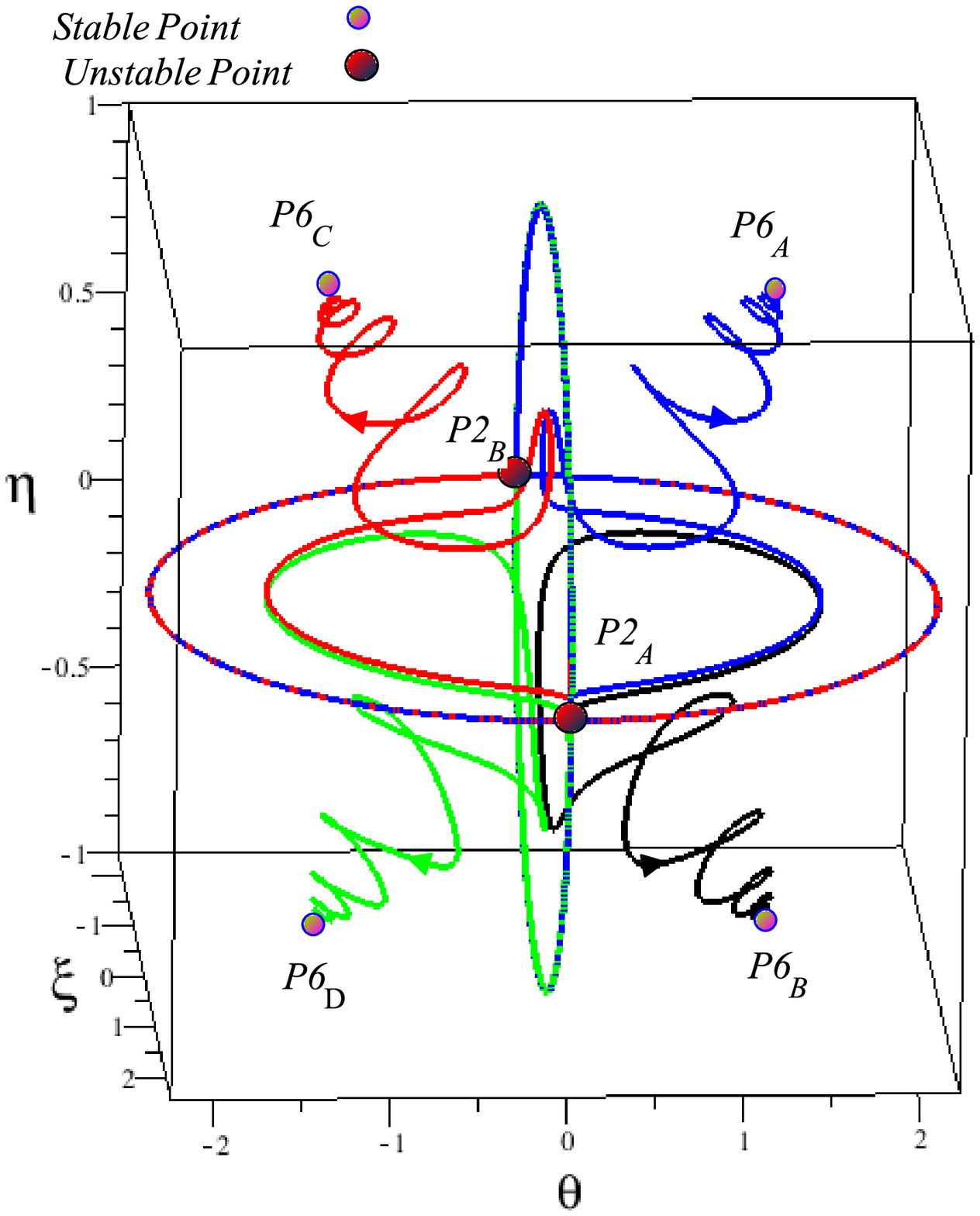}\hspace{0.1 cm}\\
Fig. 4:  The phase space graph shows the trajectories leave the unstable critical point $P2_{AB}$, \\entering the stable critical points $P6_{ABCD}$. \\
\end{tabular*}\\
In Fig. 5 we show the trajectories begins from unstable states $P2_{A}$ or $P2_{B}$, and approaches the stable states $P4_{A}$ or  $P4_{B}$. The experimentally favored one is also depicted with red color.

\begin{tabular*}{2.5 cm}{cc}
\includegraphics[scale=.4]{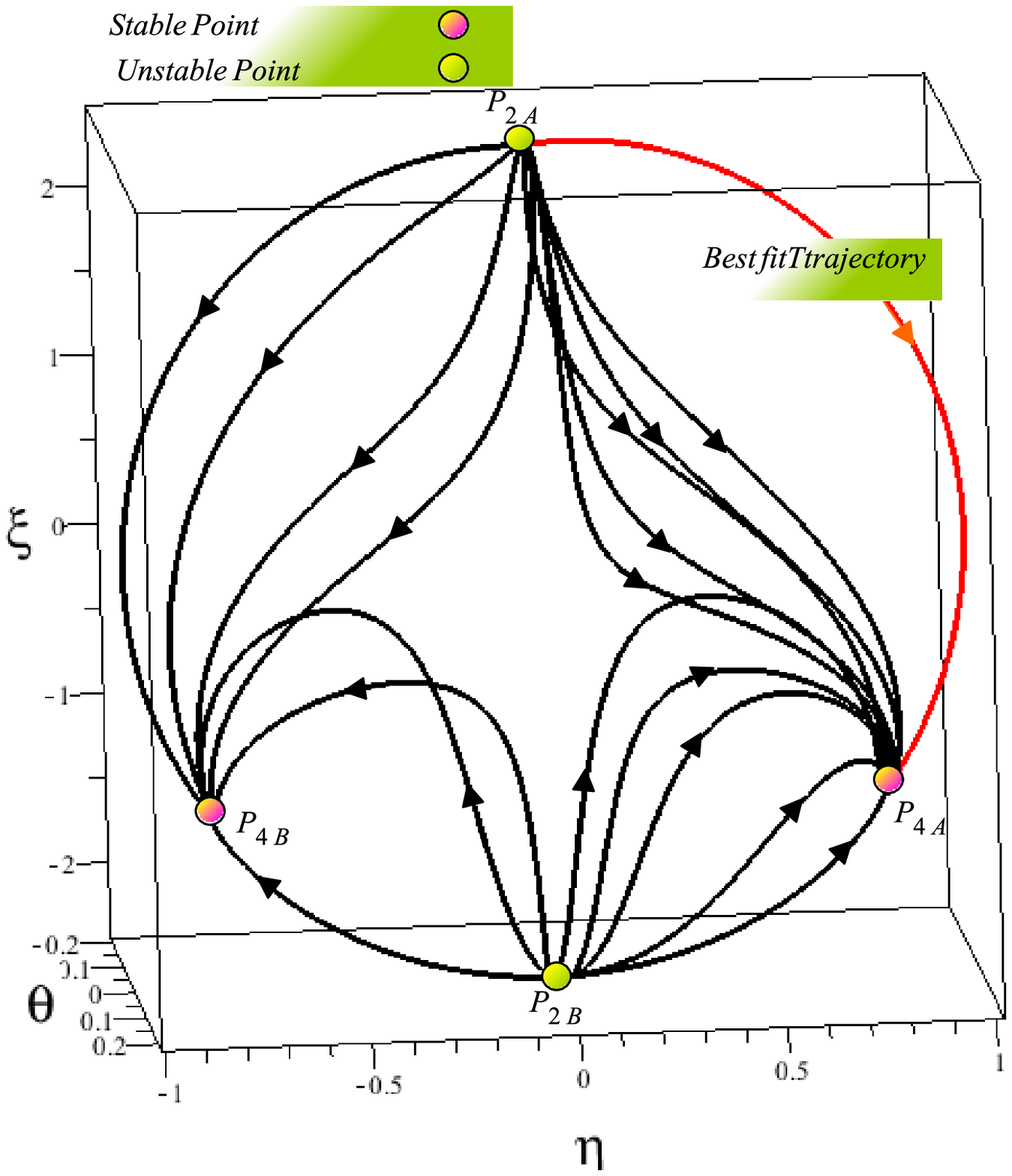}\hspace{0.1 cm}\\
Fig. 5:  The phase space graph shows the trajectories leave the unstable critical point $P2_{AB}$, \\entering the stable critical points $P4_{AB}$. The best fitted trajectory is also depicted with red color.
\end{tabular*}\\

In the next section we shall discuss the strengths and limits of the model explicitly by directly or indirectly testing it against other observational data and physical parameters.

\section{Cosmological tests}

In this section, for the model under consideration, three cosmological tests are performed to verify the model's validity. One of the most popular tests is the dynamic of the reconstructed effective EoS parameter of the model with regards to the constrained parameters. The reconstructed effective EoS parameter in terms of new dynamical variables is given by $\omega_{eff}=\frac{\omega\xi^{2}}{6}-\eta^{2}$. From numerical calculation, in Fig.6, the effective EoS parameter for both best fitted and arbitrary parameters are plotted. From the graph, all the trajectories begins from unstable critical point $P2_A$ in the past and approaches the stable critical point $P4_A$ in future. All display phantom crossing behavior in the past or future. The current best fitted effective EoS parameter is about  $\omega_{eff}=-0.9$ which is within the limit of our observations \cite{Rapetti}.\\

\begin{tabular*}{2.5 cm}{cc}
\includegraphics[scale=.5]{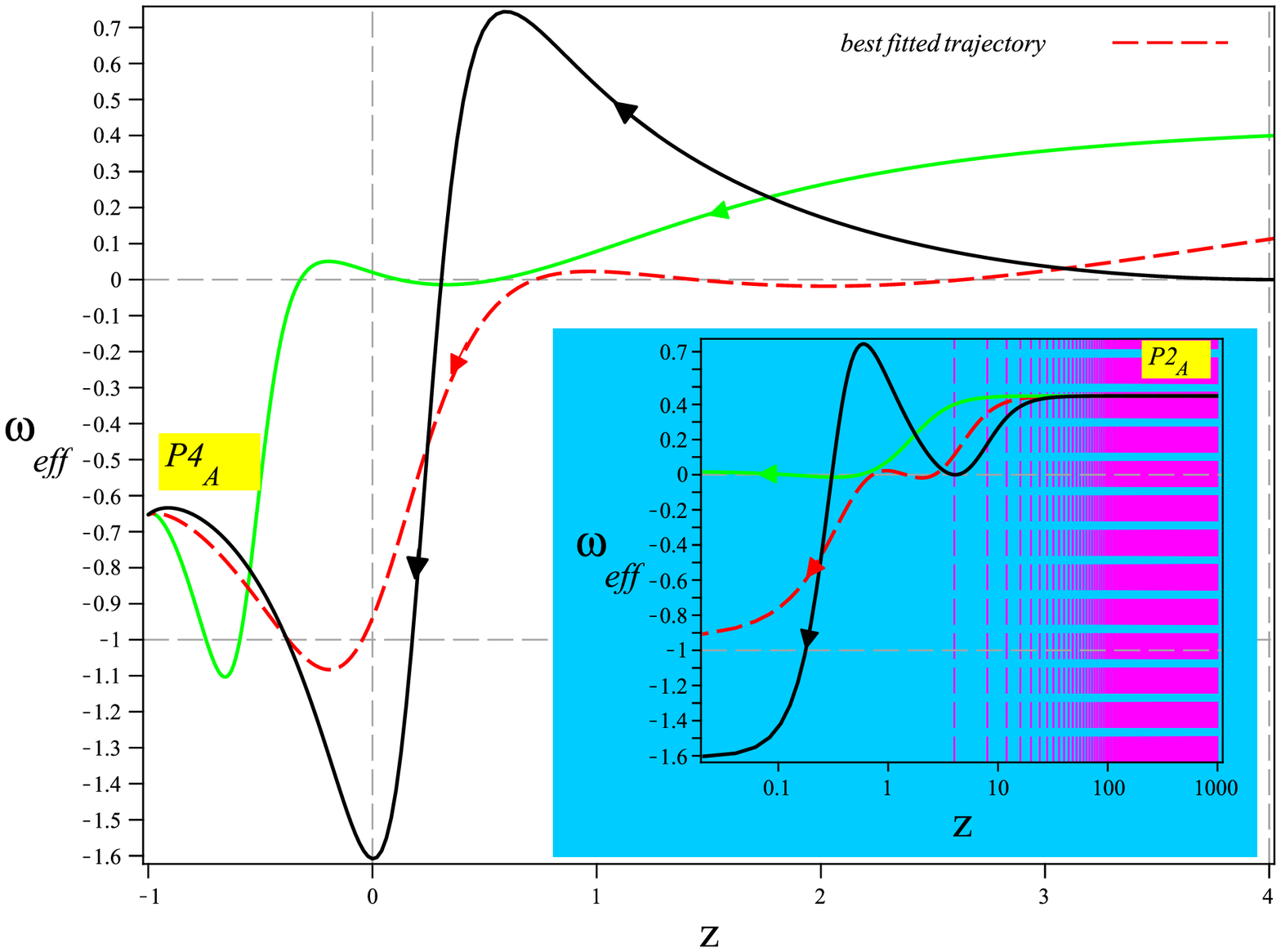}\hspace{0.1 cm}\\
Fig. 6:  The EoS parameter as a function of redshift\\
\end{tabular*}\\

In direct touch with observation, we test the dynamics of the Hubble parameter derived from numerical calculation in our model against its observational data \cite{hubbledata}. Fig. 7 reveals that only for the best fitted Hubble parameter obtained from numerical calculation is relatively in good agreement with the observational data.

\begin{tabular*}{2.5 cm}{cc}
\includegraphics[scale=.4]{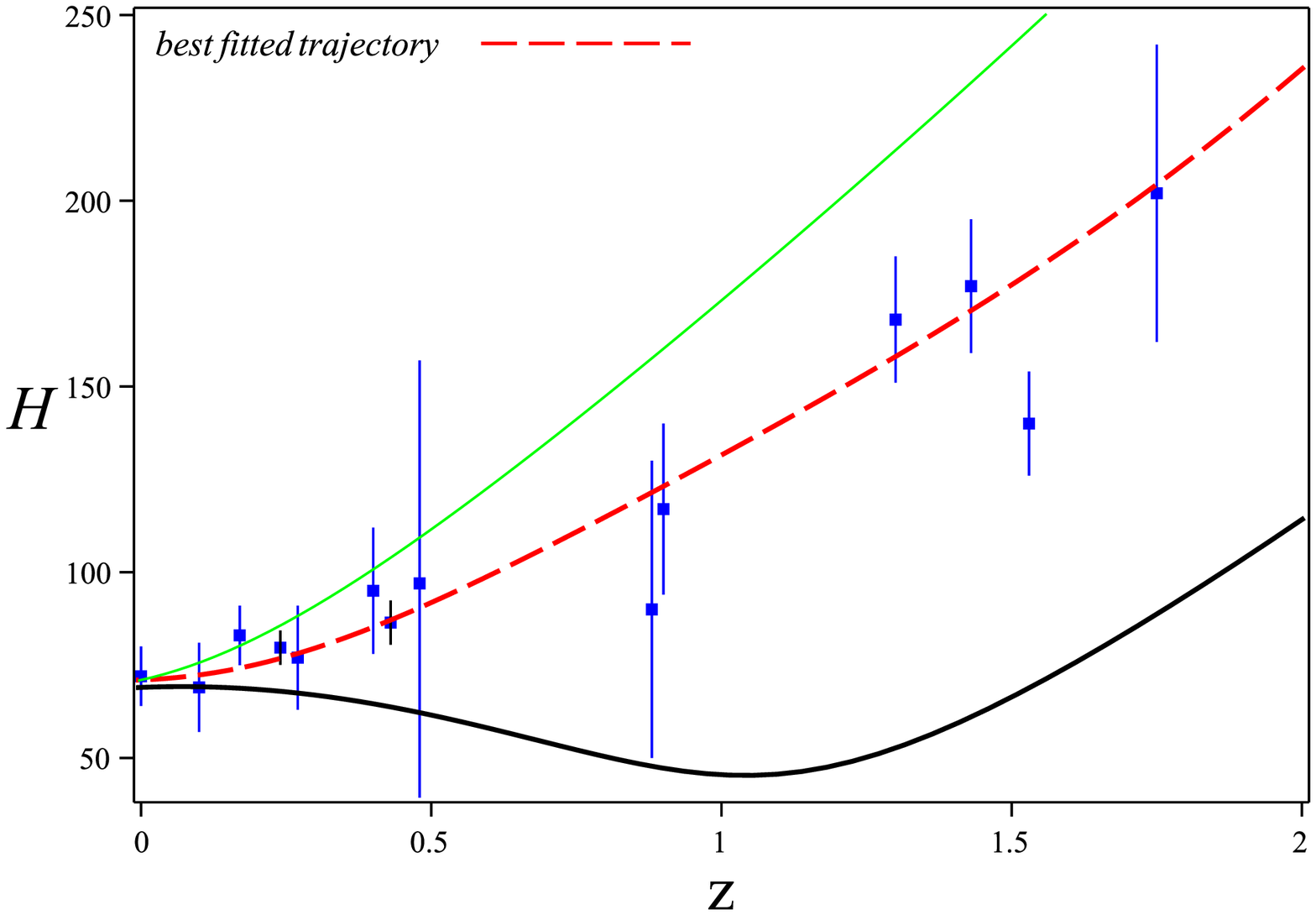}\hspace{0.1 cm}\\
Fig. 7:  The graph of hubble parameter H(z) in comparison with the observational data. \\
\end{tabular*}\\

Further, we examine the behavior of the time shift density parameter, $\Delta \alpha/\alpha$, against observational data \cite{webb5}. The Fig. 8 shows that again only the best fitted trajectory passes is in good agreement with the observation.

Though the best fitted model is not very sensitive to $\alpha$ variation in $10^{-5}$ scale, but it passes through the datapoints. From the graph, the model extrapolates that at very high resdshifts about $z\sim 1000$, the $\alpha$ variation is less than $10^{-2}$ which is consistent with the observational data from the power spectrum of anisotropy in the cosmic microwave background (CMB)  \cite{alpha-cmb}-\cite{Nollett}.

\begin{tabular*}{2.5 cm}{cc}
\includegraphics[scale=.4]{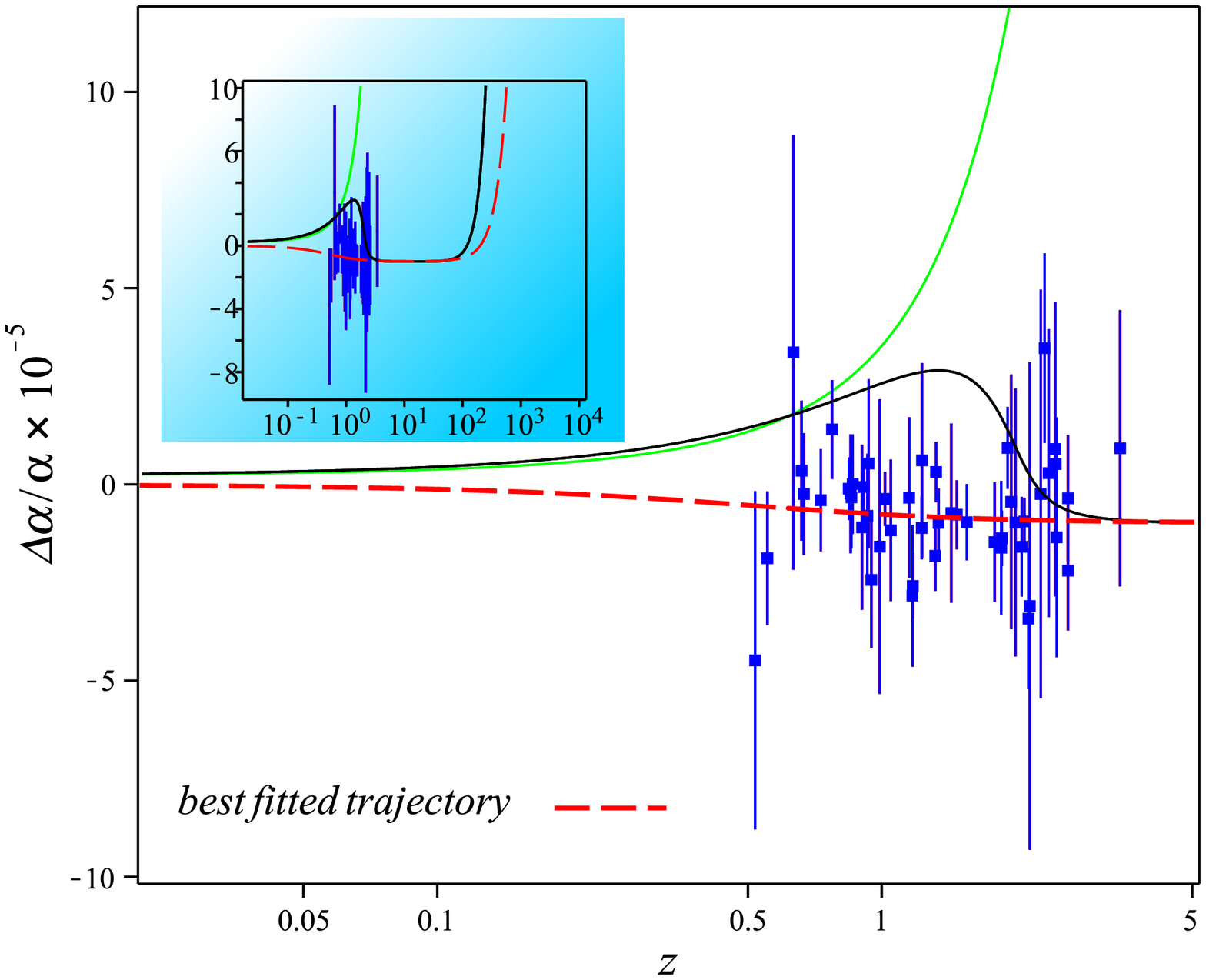}\hspace{0.1 cm}\\
Fig. 8:  The time shift density parameter, $\Delta \alpha/\alpha$, plotted for the model against\\quasar absorption spectra observations   \\
\end{tabular*}\\

From Fig. 9, the temporal drift in the value of $\alpha$, i.e. $\dot{\alpha}/\alpha$, obtained from numerical calculation is compared with the observational data for the corresponding redshifts of virialisation, $z_v$ \cite{mota-alphadotdata}. The result again for the best fitted model parameters seems reasonably supported by the data.

\begin{tabular*}{2.5 cm}{cc}
\includegraphics[scale=.4]{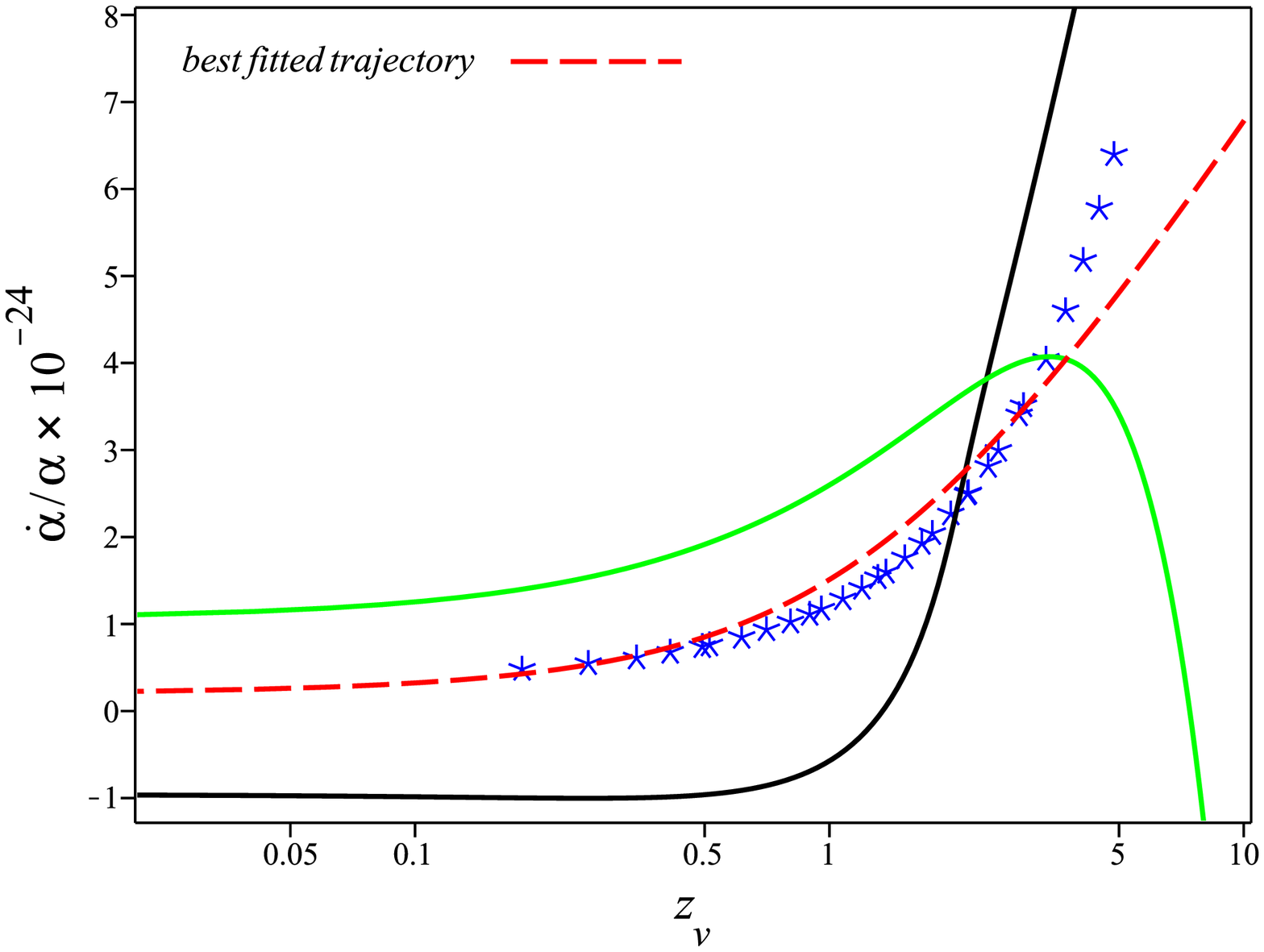}\hspace{0.1 cm}\\
Fig. 9:  The temporal drift in the value of $\alpha$, $\dot{\alpha}/\alpha$, plotted for the model\\ against observational data for the corresponding redshifts of virialisation, $z_v$.\\
\end{tabular*}\\

Finally, we reconstruct the scalar field responsible for both $\alpha$ variation and universe acceleration using the best fitted model parameters. Fig 10 shows that the best fitted scalar field variation within the range of redshift $0.1<z<10000$ is of order $10^{-5}$ as expected.

\begin{tabular*}{2.5 cm}{cc}
\includegraphics[scale=.4]{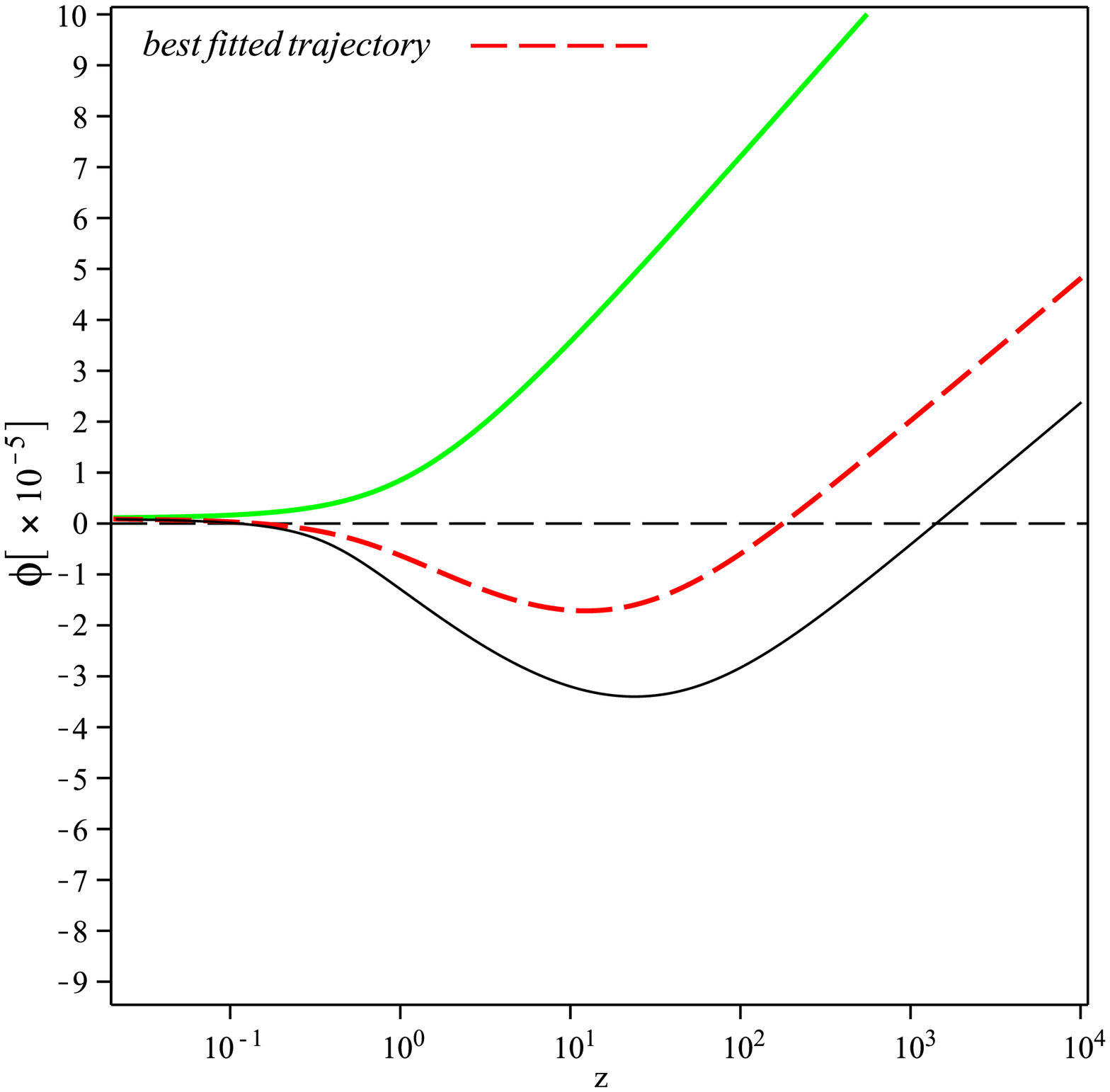}\hspace{0.1 cm}\\
Fig. 10:  The reconstructed scalar field $\phi$ as a function of the redshift\\
\end{tabular*}\\

\section{Discussion and Summary }

In this paper we have introduced a criterion that can be used to probe both the cosmological
viability of $\alpha$ variation theories and universe acceleration with the scalar field
in BSBM theory is responsible for both of them. To implement the idea, the model has
to simultaneously satisfy both observational evidence from SNe Ia and quasar absorption
spectra. The first direct evidence for cosmic acceleration comes from SNe Ia dataset that
provides the strongest constraints on the equation of stat parameter. Therefore, we first
best fit the model with these data for distance modulus and find constraints on the model
parameters and initial conditions. We also investigate the phase space of the model. Stability
analysis reveals that the best fitted model begins from an unstable state in the past
and moves towards a stable state in future. We then perform quantitative and qualitative
analysis to validate the theory and the constraints on its parameters by experiment. Two
quantitative tests are performed, the observational hubble parameter test and $\alpha$ variation.
The best fitted model is verified by these two tests. The result shows that with the best fitted
parameters the universe never underwent phantom era in the past, while crossing occurs
twice in near future. It is notable that best fitting the model parameters with the observational
data of SNe Ia and testing the model with high redshift quasar absorption spectra in
our scenario seems compatible with our results. Obviously, an improved measurements of
the redshift dependence of $\alpha$ variation will discover a better explanation to both late time
universe acceleration and "fundamental constants" variation in relation to dark energy.

\end{document}